\documentclass[prd,tightenlines,floatfix,showpacs,preprintnumbers,nofootinbib,eqsecnum,superscriptaddress,twocolumn]{revtex4-1}

\usepackage{color}      
\usepackage{amsbsy}     
\usepackage{amsfonts}   
\usepackage{amsmath}    
\usepackage{amssymb}    
\usepackage{xcolor}
\usepackage{wasysym}
\usepackage{multirow}
\usepackage{mciteplus}
\usepackage{psfrag}
\usepackage{slashed}
\usepackage{cancel}
\usepackage{graphicx}
\usepackage{ulem}

\usepackage{hyperref}

\newcommand{\DMNLO}{{\tt DM@NLO}}

\newcommand{\MSbar}{{$\overline{\mathrm{MS}}$}}
\newcommand{\DRbar}{{$\overline{\mathrm{DR}}$}}

\catcode`\@=11
\font\manfnt=manfnt
\def\Watchout{\@ifnextchar [{\W@tchout}{\W@tchout[1]}}
\def\W@tchout[#1]{{\manfnt\@tempcnta#1\relax%
  \@whilenum\@tempcnta>\z@\do{%
    \char"7F\hskip 0.3em\advance\@tempcnta\m@ne}}}
\let\foo\W@tchout
\def\dubious{\@ifnextchar[{\@dubious}{\@dubious[1]}}

\def\@dubious[#1]{%
  \color{red}\setbox\@tempboxa\hbox{\@W@tchout#1}
  \@tempdima\wd\@tempboxa
  \list{}{\leftmargin\@tempdima}\item[\hbox to 0pt{\hss\@W@tchout#1}]}
\def\@W@tchout#1{\W@tchout[#1]}
\catcode`\@=12

\begin{document}
\preprint{LAPTH-045/19, MS-TP-19-28, TUM-HEP-1226-19}

\title{SUSY-QCD corrected and Sommerfeld enhanced stau annihilation into heavy quarks with scheme and scale uncertainties}


\author{J.~Branahl}
\email{j\_bran33@uni-muenster.de}
\affiliation{
	Institut f\"ur Theoretische Physik, Westf\"alische Wilhelms-Universit\"at M\"unster, Wilhelm-Klemm-Stra{\ss}e 9, D-48149 M\"unster, Germany
}

\author{J.~Harz}
\email{julia.harz@tum.de}
\affiliation{
	Physik Department T70, James-Franck-Stra{\ss}e, Technische Universit\"at M\"unchen, D-85748 Garching, Germany
}

\author{B.~Herrmann}
\email{herrmann@lapth.cnrs.fr}
\affiliation{
	Univ.\ Grenoble Alpes, Univ.\ Savoie Mont Blanc, CNRS, LAPTh, F-74000 Annecy, France
}

\author{M.~Klasen}
\email{michael.klasen@uni-muenster.de}
\affiliation{
	Institut f\"ur Theoretische Physik, Westf\"alische Wilhelms-Universit\"at M\"unster, Wilhelm-Klemm-Stra{\ss}e 9, D-48149 M\"unster, Germany
}

\author{K.~Kova\v{r}\'ik}
\email{karol.kovarik@uni-muenster.de}
\affiliation{
	Institut f\"ur Theoretische Physik, Westf\"alische Wilhelms-Universit\"at M\"unster, Wilhelm-Klemm-Stra{\ss}e 9, D-48149 M\"unster, Germany
}
 
\author{S.~Schmiemann}
\affiliation{
	Institut f\"ur Theoretische Physik, Westf\"alische Wilhelms-Universit\"at M\"unster, Wilhelm-Klemm-Stra{\ss}e 9, D-48149 M\"unster, Germany
}

\date{\today}

\begin{abstract}
We investigate stau-antistau annihilation into heavy quarks in the phenomenological Minimal Supersymmetric Standard Model within the \DMNLO\  project. We present the calculation of the corresponding cross section including corrections up to $\mathcal{O}(\alpha_s)$ and QED Sommerfeld enhancement. The numerical impact of these corrections is discussed for the cross section and the dark matter relic density, where we focus on top-quark final states and consider either neutralino or gravitino dark matter. Similarly to previous work, we find that the presented corrections should be included when calculating the relic density or extracting parameters from cosmological observations. Considering scheme and scale variations, we estimate the theoretical uncertainty that affects the prediction of the annihilation cross section and thus the prediction of the relic density.
\end{abstract}


\maketitle

\section{Introduction}
\label{Sec:Intro}

More than 80 years after its first observation \cite{Zwicky}, the existence of dark matter in our Universe is now well established by various coinciding observations (see review \cite{Freese2017} and references therein). In the absence of a clear consensus about the exact nature of dark matter, numerous theoretical models have been developed to explain its nature. Most of such models are based on the hypothesis that dark matter consists of weakly interacting massive particles (WIMPs) which achieve the observed relic abundance through thermal freeze-out \cite{KolbTurner, Klasen:2015uma}.

In the present work, we focus on the Minimal Supersymmetric Standard Model (MSSM), which provides suitable WIMP candidates \cite{Ellis1983} such as the lightest neutralino or the gravitino. Assuming that $R$-parity is conserved, these particles are stable, and they interact only weakly as required by the WIMP paradigm.

Over the last decades, the relic abundance of cold dark matter (CDM) within the cosmological $\Lambda$CDM model has been determined to a very good precision,
\begin{equation}
	\Omega_{\mathrm{CDM}}h^2 = 0.1200 \pm 0.0012 \,,
	\label{Eq:Planck}
\end{equation}
where $h$ denotes the present Hubble expansion rate in units of 100 km s$^{-1}$ Mpc$^{-1}$. This interval has been obtained from the cosmic microwave background measurements by the Planck satellite \cite{Planck} combined with polarization data from the WMAP mission \cite{WMAP9}. Within a given particle physics model, such as the MSSM, the relic density of dark matter can be theoretically predicted, allowing to identify cosmologically favoured regions of parameter space. More precisely, for a dark matter candidate $\chi$ with mass $m_{\chi}$, the predicted relic density $\Omega_{\chi}h^2$ is obtained through
\begin{equation}
	\Omega_{\chi} ~=~ \frac{m_{\chi} n_{\chi}}{\rho_{\rm crit}} \,.
\end{equation}
Here, $\rho_{\rm crit}$ stands for the critical energy density of the Universe and $n_{\chi}$ for today's number density of the dark matter candidate. The value of $n_{\chi}$ corresponds to the solution of the  Boltzmann equation \cite{Gondolo1990, Griest1990, Edsjo1997}
\begin{equation}
	\frac{\mathrm{d}n_\chi}{\mathrm{d}t} = -3 H n_\chi 
		- \left\langle\sigma_{\mathrm{ann}}v\right\rangle \Big[ n_\chi^2 
		- \left( n_\chi^{\mathrm{eq}} \right)^2 \Big] \,.
	\label{Eq:Boltzmann1}
\end{equation}
This differential equation describing the time evolution of the number density contains the thermally averaged cross section $\langle \sigma_{\rm ann}v \rangle$ of the annihilating neutralinos.

Dark matter annihilation cross sections typically lead to a relic density exceeding the limits given in Eq.\ \eqref{Eq:Planck} causing the overclosure of the Universe. As a consequence, the annihilation cross section needs to be enhanced by some mechanism, such as resonant annihilation or efficient co-annihilations, resulting in lower values of the relic density. In the present paper, we will focus on the latter case, in particular we will assume that the lightest neutralino and the lightest stau are almost degenerate in mass.

Accounting for co-annihilations, the thermally averaged annihilation cross section appearing in Eq.\ \eqref{Eq:Boltzmann1} can be expressed as 
\begin{align}
    \langle \sigma_{\rm ann} v \rangle ~=~ 
    \sum_{i,j} \langle \sigma_{ij} v_{ij} \rangle 
    \frac{n_i^{\rm eq}}{n_{\chi}^{\rm eq}} \frac{n_j^{\rm eq}}{n_{\chi}^{\rm eq}} \,,
    \label{Eq:CrossSection}
\end{align}
where the relative velocities are given by $v_{ij} = \sqrt{(p_i\cdot p_j)^2}-m_i^2m_j^2/(E_i E_j)$. The ratios of equilibrium densities appearing in Eq.\ \eqref{Eq:CrossSection} are suppressed via the so-called Boltzmann factors
\begin{align}
    \frac{n_i^{\rm eq}}{n_{\chi}^{\rm eq}} ~\propto~
        \exp{\left\{-\frac{m_i-m_{\chi}}{T}\right\}} \,.
    \label{Eq:ExpFactor}
\end{align}
This implies that only co-annihilation of dark matter with particles that are almost degenerate in mass can have a sizeable impact on the relic density. In the present paper, we assume that the lightest neutralino and the lighter stau are very close in mass. 

The very narrow observational interval given in Eq.\ \eqref{Eq:Planck} clearly calls for very precise theoretical predictions. Within public codes calculating the dark matter relic density for new physics models, such as {\tt micrOMEGAs} \cite{MO2001, MO2006, MO2013, MO2016, MO2018} or {\tt DarkSUSY} \cite{DS2000, DS2004, DS2018}, the underlying processes entering Eq.\ \eqref{Eq:CrossSection} are evaluated usually at tree-level, including effective couplings capturing certain higher order effects, e.g., for the Yukawa couplings. It is the goal of the {\tt DM@NLO} project to provide a more accurate calculation of the annihilation cross section and thus of the dark matter relic density.

Over the last decade, we have demonstrated the impact of QCD next-to-leading order (NLO) corrections in the following cases within the MSSM: Gaugino annihilation into quarks \cite{AFunnel, ChiChi2qq1, ChiChi2qq2, ChiChi2qq3}, neutralino-stop co-annihilation into several final states \cite{Freitas2007, NeuQ2qx1, NeuQ2qx2}, stop-antistop annihilation into electroweak final states \cite{StSt2xx}, and stop-stop annihilation into quarks \cite{StSt2qq}. Moreover, we have been able to evaluate the theoretical uncertainty of the relic density calculation \cite{scalevar}. Other authors have shown that electroweak corrections may have an equally sizeable impact \cite{Boudjema2005, Baro2007, Baro2009}, and the impact of Sommerfeld enhancement has been studied in various cases \cite{Slatyer:2009vg, Drees:2009gt, Feng:2010zp, Hryczuk:2011vi, Hryczuk:2011tq, Beneke2014a, Beneke2014b}. 

The present paper will add to the above list of processes by presenting the corrections of order $\alpha_s$ to stau-antistau annihilation. This process may be relevant in scenarios with neutralino or gravitino dark matter. In the following, we start by discussing the phenomenological impact of stau-antistau annihilation in Sec.\ \ref{Sec:Pheno}. In Sec.\ \ref{Sec:Calculation}, we will first discuss the technical details of the NLO calculation. Moreover, we will present the QED Sommerfeld enhancement included in our calculation. In Sec.\ \ref{Sec:Results}, we will illustrate the effect of the corrections in typical scenarios within the phenomenological MSSM for both neutralino and gravitino dark matter. We will also discuss the theoretical uncertainty coming from the variations of the renormalization scale and the renormalization scheme. Conclusions are given in Sec.\ \ref{Sec:Conclusions}. 
\section{Phenomenology related to stau-antistau annihilation}
\label{Sec:Pheno}

Let us start by discussing the phenomenology of stau-antistau annihilation in the context of the dark matter relic density. This process may become relevant in two cases. 

First, as mentioned in the introduction, if the stau is very close in mass to the neutralino, neutralino-stau co-annihilation as well as stau pair-annihilation will contribute in a sizeable manner to the total annihilation cross section $\sigma_{\rm ann}$. In this case, the prediction of the neutralino relic density is obtained directly from solving the Boltzmann equation as explained in the introduction. 

The second situation is the case where the dark matter candidate is the gravitino, denoted as $\tilde{G}$, which is the spin-3/2 superpartner of the graviton, if local supergravity is assumed. In this situation, the next-to-lightest supersymmetric particle may be either the lightest gaugino or the lightest sfermion, for example the lighter stau. The phenomenology related to the additional gravitino is governed by a single additional parameter, which is the gravitino mass $m_{\tilde{G}}$. Recent detailed discussions of gravitino dark matter within the phenomenological Minimal Supersymmetric Standard Model (pMSSM) can be found, among others, in Refs.\ \cite{RizzoHewett2012, Heisig2013, Arbey2015}.

In this situation, the gravitino relic density receives contributions from thermal and non-thermal production, 
\begin{align}
    \Omega_{\tilde{G}}h^2 ~=~ \Omega_{\tilde{G}}^{\rm th}h^2 + \Omega_{\tilde{G}}^{\rm non-th}h^2 \,.   
\end{align}
The thermal contribution depends on the reheating temperature, $T_R$, and the gluino mass, $m_{\tilde{g}}$, and can be approximated as \cite{Bolz2000}
\begin{align}  
    \Omega_{\tilde{G}}^{\rm th}h^2 \simeq
        0.27 \biggl( \frac{T_{\rm R}}{10^{10} \, \mathrm{GeV}} \biggl ) \! \biggl (\frac{100 \, \mathrm{GeV}}{m_{\tilde{G}}} \biggl ) \! \biggl (\frac{m_{\tilde{g}}}{1 \, \mathrm{TeV}} \biggl )^{\!\! 2} .
    \label{Eq:Omh2GravTh}
\end{align} 
The corresponding full expression has been derived in Refs.\ \cite{Pradler2006b, Rychkov2007}. In the present work we focus on cases where the non-thermal contribution dominates, such that it is reasonable to rely on the simplified expression given in Eq.\ \eqref{Eq:Omh2GravTh}.

The non-thermal contribution arises from the decay of the lighter stau. If $R$-parity is conserved, each stau can decay only into a gravitino, and the corresponding contribution to the gravitino relic density is obtained by reweighing the would-be relic density of the stau, obtained from integrating the Boltzmann equation, according to \cite{Pradler2006a}
\begin{align} 
    \Omega_{\tilde{G}}^{\rm non-th}h^2 ~=~ 
        \frac{m_{\tilde{G}}}{m_{\tilde{\tau}_1}} \, \Omega^{\rm th}_{_{\tilde{\tau}_1}}h^2 \,. 
    \label{Eq:Omh2GravNonTh}
\end{align}
In this context, the stau lifetime is constrained in order to preserve the abundances of light elements in the early Universe, which are well explained by primordial nucleosynthesis. In terms of stau and gravitino masses, this constraint can be approximated as \cite{Pospelov2008, Herrmann2008}
\begin{align}
    t_{\tilde{\tau}_1} \simeq \bigl( 6100\,{\rm s}\bigr) \left( \frac{1~{\rm TeV}}{m_{\tilde{\tau}_1}} \right)^{\!\! 5} \left( \frac{m_{\tilde{G}}}{100~{\rm GeV}} \right)^{\!\! 2} \lesssim 6000\,{\rm s}\,,
    \label{Eq:StauLifetime}
\end{align}
implying that the gravitino mass is about one order of magnitude smaller than the stau mass.

In the following, we illustrate the phenomenology, and later on also the impact of higher-order corrections to the stau annihilation cross section, for two example scenarios within the pMSSM, where the 19 independent soft-breaking parameters are defined at the supersymmetry (SUSY) scale $Q = \sqrt{m_{\tilde{t}_1} m_{\tilde{t}_2}}$. In order to identify representative parameter configurations, we make use of the pMSSM analysis presented by ATLAS in Ref.\ \cite{ATLAS2015pMSSM}. The parameter points resulting from this study satisfy the imposed constraints from LHC searches, the observed Higgs mass of $m_{h^0} \sim 125$ GeV \cite{ATLASHiggs, CMSHiggs, ATLASCMSHiggs}, and rare decays such as, e.g., $b \to s\gamma$ and $B_s \to \mu^+\mu^-$. Let us also stress that the corresponding parameter region is robust against more recent searches performed by the ATLAS and CMS collaborations \cite{CMS:2019eln}.

Inspired from the parameter points given in Ref.\ \cite{ATLAS2015pMSSM}, we have chosen a typical scenario for our study related to neutralino dark matter. All pMSSM parameters related to this scenario, labelled I, are given in Tab.\ \ref{Tab:Scenarios}, together with the corresponding neutralino and stau masses, which are relevant in our study.

As the ATLAS analysis only covers the case of neutralino dark matter, we have defined a second scenario, labelled II, for our study of gravitino dark matter. This second scenario aims at illustrating the situation in a rather simple way, the soft parameters being chosen such that only the staus are relatively light with masses just below 2 TeV, while all other states are rather heavy with masses of about 5 TeV. The corresponding soft-breaking parameters, together with the relevant masses, are displayed in Tab.\ \ref{Tab:Scenarios}. In addition, in order to satisfy the lifetime constraint mentioned above, the gravitino mass will be chosen to be around 400 GeV. This implies a reheating temperature of $T_{\mathrm{R}} \approx \mathcal{O}(10^{7})~\mathrm{GeV}$ in order to meet the observed relic density in Eq.~\eqref{Eq:Planck}.

\begin{table}
	\centering
	\begin{tabular}{|c|cc|}
	    \hline
		 & I & II \\
		\hline
		\hline
	    $M_{\tilde{q}_L}$ &  ~1599.9 ~ & ~5000~ \\
	    $M_{\tilde{t}_L}$ &  ~3007.0~ & ~5000~ \\
	    $M_{\tilde{u}_R}$ &  ~3904.4~ & ~5000~ \\
	    $M_{\tilde{t}_R}$ &  ~3093.0~ & ~5000~ \\
	    $M_{\tilde{d}_R}$ &  ~3096.7~ & ~5000~ \\
	    $M_{\tilde{b}_R}$ &  ~581.6~ & ~5000~ \\
		\hline
	    $M_{\tilde{\ell}_L}$ &  ~3586.7~ & ~5000~ \\
	    $M_{\tilde{\tau}_L}$ & ~563.6~ & ~1800~ \\
	    $M_{\tilde{\ell}_R}$ &  ~3950.4~ & ~5000~ \\
	    $M_{\tilde{\tau}_R}$ &  ~585.5~ & ~1846~ \\	\hline
	    \hline
	    $Q$ &  ~3047.8~ & ~5000~ \\
	    \hline
	\end{tabular}
    \qquad
	\begin{tabular}{|c|cc|}
	    \hline
	  & I & II \\
		\hline
		\hline
		$M_1$ & ~546.0~ & ~5000~\\
	    $M_2$ &  ~-3461.7~ & ~5000~\\
	    $M_3$ &  ~3126.7~ & ~5000~ \\
		\hline
	    $A_t$ &  ~5246.7~ & ~-3000~ \\
	    $A_b$ & ~-2530.3~ & ~1000~ \\
	    $A_{\tau} $ &  ~1586.4~ & ~5000~ \\
		\hline
		~$\tan\beta$~ & ~18.0~ & ~22.0~ \\
	    $\mu$ &  ~2643.6~ & ~5000~ \\
	    $m_{A^0}$ &  ~2962.3~ & ~5000~ \\
	    \hline
	    \hline
	    $m_{\tilde{\chi}^0_1}$ & ~540.6~ & ~4915.8~ \\
	    $m_{\tilde{\tau}_1}$ & 540.7 & 1810.8 \\
	    \hline
    \end{tabular}
	\caption{Scalar soft mass parameters, gaugino mass parameters, trilinear couplings, and parameters related to the Higgs sector at the input scale $Q$ for two reference scenarios I and II within the pMSSM. We also indicate the resulting physical masses of the lightest neutralino and the lighter stau. The values of the remaining physical masses are not displayed here, as they are not relevant for our study. The gravitino mass for the study of Scenario II will be specified in Sec.\ \ref{Sec:RelicIb}. All dimensionful quantities are given in GeV.}
	\label{Tab:Scenarios}
\end{table}

For each parameter point, the physical mass has been computed from the soft-breaking parameters using {\tt SPheno 3.3.3} \cite{SPheno2003, SPheno2012}. In both scenarios, the lightest neutralino is a pure bino, while the lighter stau is strongly mixed, the mixing angle corresponding to $\cos^2\theta_{\tilde{\tau}} \approx 0.42$ and $\sin^2\theta_{\tilde{\tau}} \approx 0.58$ for Scenario I and $\cos^2\theta_{\tilde{\tau}} \approx \sin^2\theta_{\tilde{\tau}} \approx 0.50$ for Scenario II.
 
Coming to the calculation of the relic density, the physical mass spectrum obtained from {\tt SPheno} is handed over to {\tt micrOMEGAs 2.4.1} \cite{MO2006} using the {\it SUSY Les Houches Accord 2} format \cite{SLHA1, SLHA2}. {\tt micrOMEGAs} then performs the numerical integration of the Boltzmann equation based on the annihilation cross section computed by {\tt CalcHEP} \cite{CalcHEP}. We will use {\tt micrOMEGAs} to compute the neutralino or stau relic density, respectively. For Scenario II, the gravitino relic density will then be obtained from Eqs.\  \eqref{Eq:Omh2GravTh} and \eqref{Eq:Omh2GravNonTh} once the gravitino mass and the reheating temperature have been fixed.

\begin{table}
	\centering
	\begin{tabular}{|c|cc|}
	    \hline
		Processes & I & II \\
		\hline
		\hline
        $\tilde \tau_1 \tilde \tau_1^{*} \to t \bar{t}$ & ~~31.5~~ & ~~25.9~~ \\
        $\tilde \tau_1 \tilde \tau_1^{*} \to \gamma \gamma$ & ~12.9~ & ~21.4~ \\
        $\tilde \tau_1 \tilde \tau_1^{*} \to h^0 h^0$ & ~10.0~ & ~2.2~ \\
        $\tilde \tau_1 \tilde \chi_1^{0} \to \ell h^0$ & ~9.2~ & ~-~ \\
        $\tilde \tau_1 \tilde \tau_1^{*} \to \ell \bar{\ell}, \nu \bar{\nu}$ & ~7.4~ & ~8.4~ \\
        $\tilde \tau_1 \tilde \chi_1^{0} \to \ell Z^0$ & ~7.0~ & ~-~ \\
        $\tilde \tau_1 \tilde \chi_1^{0} \to \ell \gamma $ & ~6.0~ & ~-~ \\ 
        $\tilde \tau_1 \tilde \tau_1^{*} \to W^+ W^-$ & ~6.5~ & ~11.3~ \\
	    \hline
    \end{tabular}
	\caption{Relative contributions in percent of the dominant annihilation channels contributing to the annihilation cross section $\sigma_{\rm ann}$ in the two reference scenarios I and II defined in Tab.\ \ref{Tab:Scenarios}. Here, $\ell$ and $\nu$ denote arbitrary lepton and neutrino states, $\ell = e,\mu,\tau$ and $\nu = \nu_e, \nu_{\mu}, \nu_{\tau}$. Further contributions below 5\% are omitted.}
	\label{Tab:Channels}
\end{table}

In Tab.\ \ref{Tab:Channels}, we summarize the dominant annihilation and co-annihilation channels contributing to the total annihilation cross section $\sigma_{\rm ann}$ entering the Boltzmann equation, Eq.~\eqref{Eq:Boltzmann1}. For both parameter configurations, stau-antistau annihilation into top-antitop pairs is the dominant contribution, followed by channels which, at the one-loop level, are insensitive to QCD corrections, such as processes including neutralinos, Higgs and gauge bosons, photons, leptons, and neutrinos.\footnote{Note that in the constrained MSSM
stau-coannnihilation leads to the correct relic density only for neutralions lighter than 600 GeV \cite{Ellis:1998kh,Ellis:1999mm}. This restriction is lifted in the pMSSM, where other channels can also contribute.} 

Generally, top quarks are more important than other quarks in the final state due to the important top-quark Yukawa coupling, which is additionally $\tan\beta$ enhanced in exchanges of scalar Higgs bosons, $h^0$ and $H^0$. Note that, since only the lighter stau is relevant in the given context, the exchange of a pseudoscalar Higgs $A^0$ is absent. We therefore focus on the annihilation into top quarks, i.e.\ we consider the process $\tilde \tau_1 \tilde \tau_1^{*} \to t \bar{t}$. Providing QCD corrections to this channel means that we can correct about 32\% and 26\% of the total annihilation cross section for Scenario I and II, respectively.

\begin{figure}
    \centering
    \includegraphics[width=0.52\textwidth, trim={0.3cm 0.3cm 0 1.0cm}, clip]{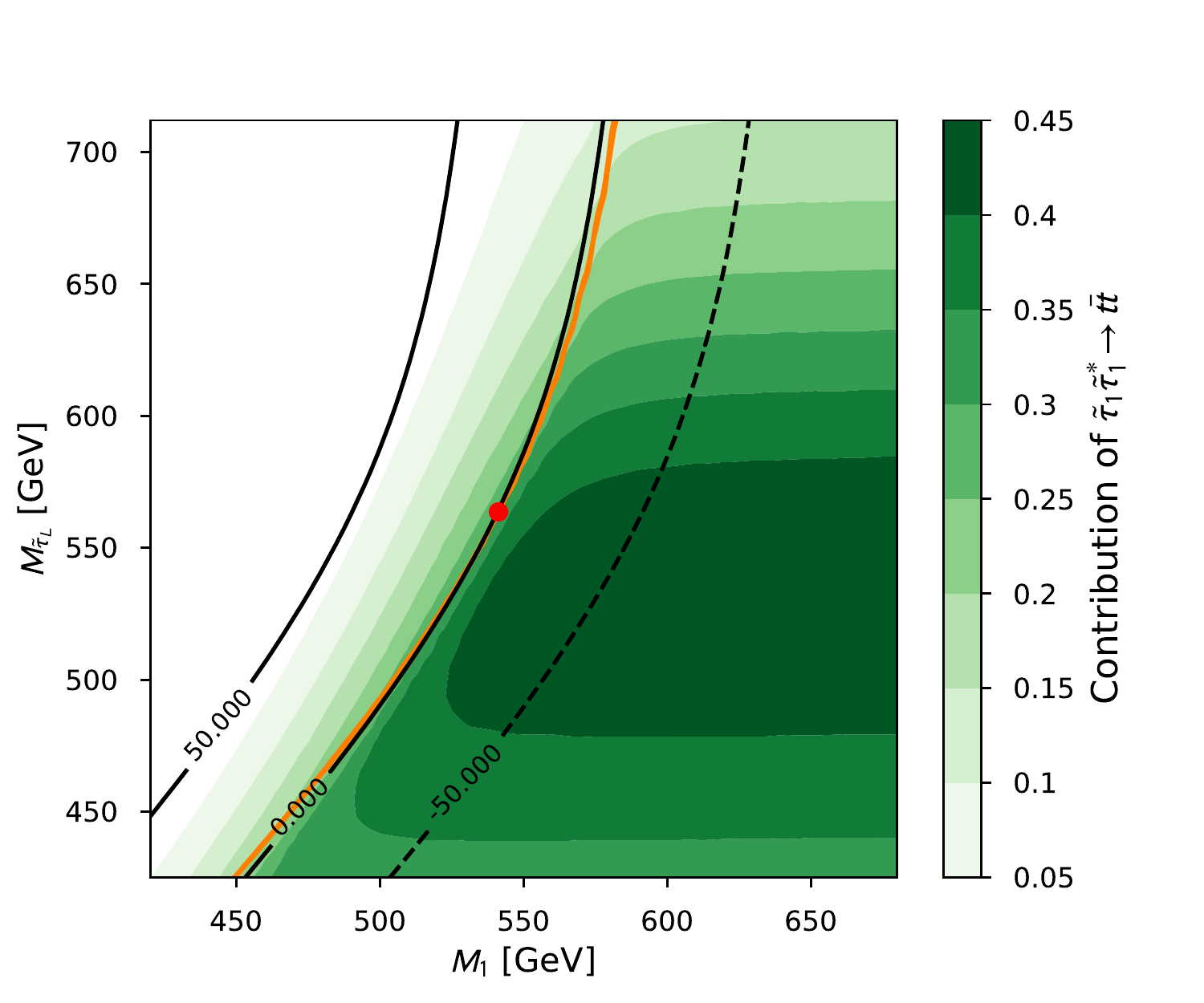}
    \caption{Parameter regions in the $M_1$--$M_{\tilde{\tau}_L}$ plane that are compatible with the Planck limits given in Eq.\ \eqref{Eq:Planck}, where the relic density has been computed using {\tt micrOMEGAs}. All other parameters are fixed to those given in Tab.\ \ref{Tab:Scenarios}. The red dot indicates Scenarios I defined in Tab.\ \ref{Tab:Scenarios}. The green contours correspond to the contribution of the process $\tilde{\tau}_1 \tilde{\tau}_1^* \to t\bar{t}$. The black contour lines indicate the difference $m_{\tilde{\tau}_1} - m_{\tilde{\chi}^0_1}$ in GeV between the physical masses of the lighter stau and the lightest neutralino.}
    \label{Fig:Omh2}
\end{figure}

We conclude the phenomenological discussion by illustrating the situation in the vicinity of our chosen reference scenario I. To this end, we show in Fig.\ \ref{Fig:Omh2} the regions corresponding to a relic density compatible with the range given in Eq.\ \eqref{Eq:Planck} obtained from the variation of the bino mass parameter $M_1$ and the stau mass parameter $M_{\tilde{\tau}_L}$ around the values given in Table \ref{Tab:Scenarios}. We also indicate the mass difference between the neutralino and the stau, and the relative contribution of the stau-antistau annihilation into top-antitop pairs which dominates this parameter region and will therefore be in the focus of our study.

As can be seen in Fig.\ \ref{Fig:Omh2}, the parameter region where the relic density agrees with the limits of Eq.\ \eqref{Eq:Planck} closely follows the line where the neutralino and the stau are equal in mass. This illustrates the importance of co-annihilations in order to obtain the observed relic density. In our Scenario I, indicated in Fig.\ \ref{Fig:Omh2} by the red dot, the small mass difference enhances the importance of the stau annihilations through the exponential factor given in Eq.\ \eqref{Eq:ExpFactor} and enables the neutralino relic density to be within the given limits.

In the remainder of this paper, we will present a higher-order calculation of the stau-antistau annihilation into top quarks and show the impact on the phenomenology discussed here.
\section{Calculation details}
\label{Sec:Calculation}

\begin{figure}
    \centering
    \includegraphics[width=0.5\textwidth]{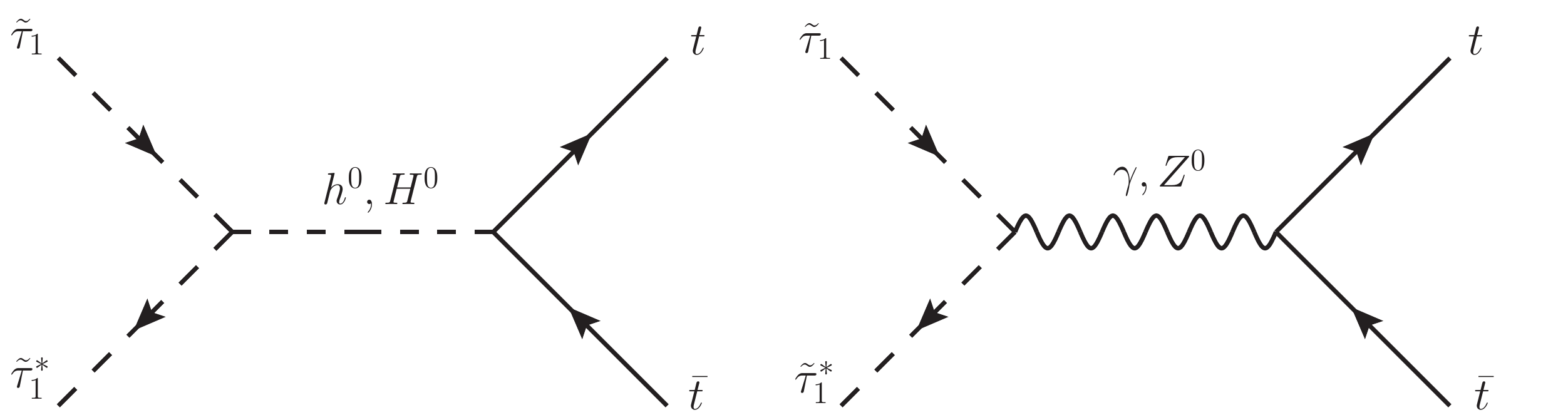}
    \caption{Tree-level Feynman diagrams for stau-antistau annihilation into top-antitop pairs via Higgs ($h^0$ or $H^0$) and vector boson ($\gamma$ or $Z^0$) exchange.}
    \label{fig:tree}
\end{figure}

In the present work, we focus on the annihilation of a stau-antistau pair into a top-antitop pair. At the tree-level, this process proceeds through the exchange of $CP$-even Higgs-boson ($h^0$ or $H^0$), a $Z^0$-boson, or a photon in the $s$-channel, as shown in Fig.\ \ref{fig:tree}. Due to the specific structure of the associated coupling with sfermions, the exchange of a pseudoscalar Higgs boson $A^0$ does not contribute in the present case of two identical stau mass eigenstates in the initial state.

In the following we will discuss higher-order corrections to the diagrams shown in Fig.\ \ref{fig:tree}. We will review the virtual and the real $\mathcal{O}(\alpha_s)$-corrections, as well as the Sommerfeld enhancement due to multiple-photon exchanges between the initial state particles.

\subsection{NLO corrections}
\label{Sec:NLO}

\begin{figure}
    \centering
    \includegraphics[width=0.45\textwidth]{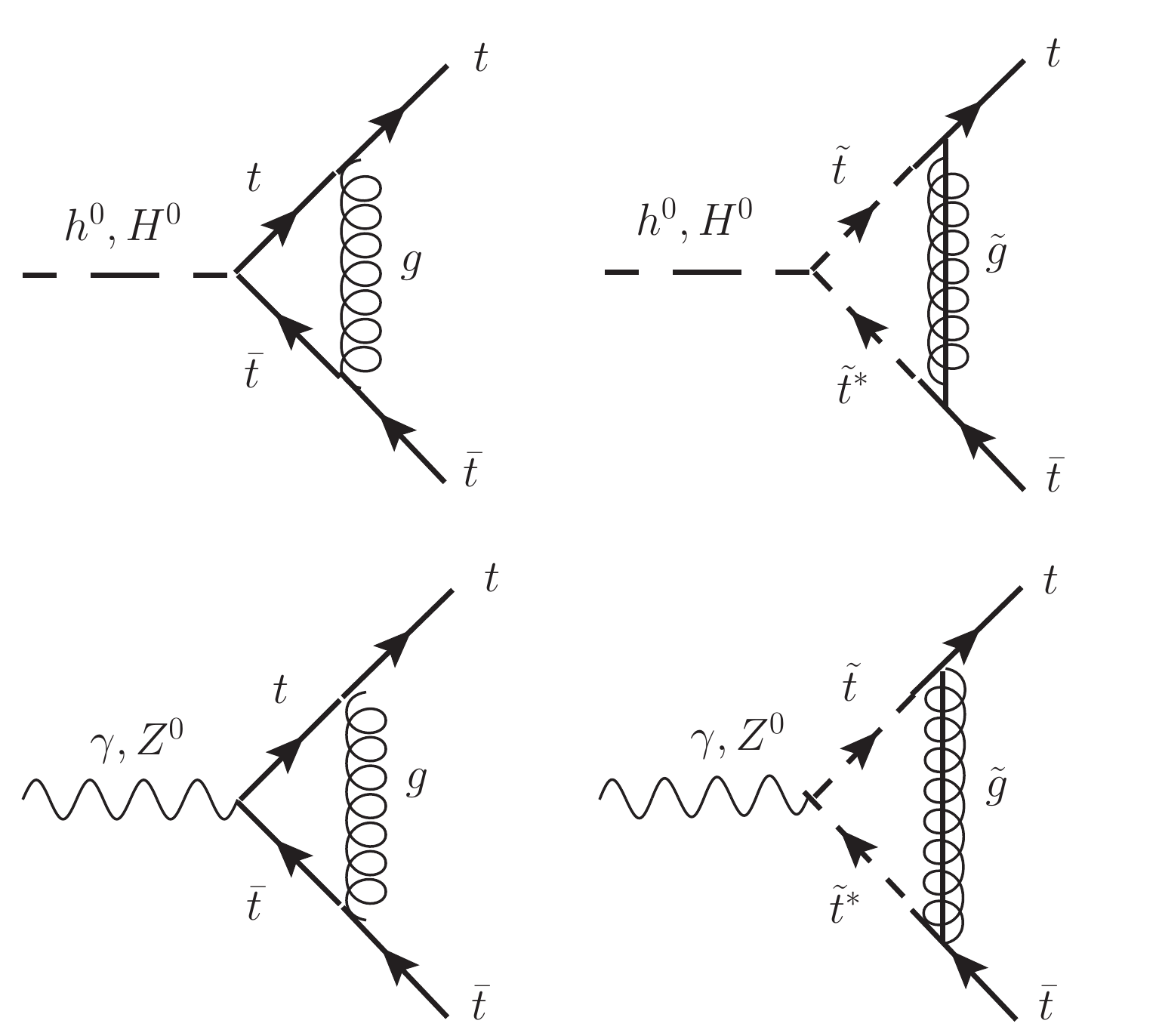}
    \caption{QCD one-loop corrections to the Higgs-top-top and vector-top-top vertices appearing in the processes of Fig.\ \ref{fig:tree}.}
    \label{fig:virtual}
\end{figure}

Virtual corrections proportional to the strong coupling constant $\alpha_s$ only arise for the final state vertex through the exchange of a gluon or a gluino between the quarks as shown in Fig.\ \ref{fig:virtual}. In order to regulate the arising divergences, we have evaluated the loop integrals \cite{PV1978} appearing in the vertex diagrams in $D=4-2\epsilon$ dimensions using the dimensional reduction (\DRbar) scheme, which preserves supersymmetry. The ultraviolet divergences introduced in the loop integrals are cancelled by a renormalization of the model parameters and fields. 

\begin{figure}
    \centering
    \includegraphics[width=0.50\textwidth]{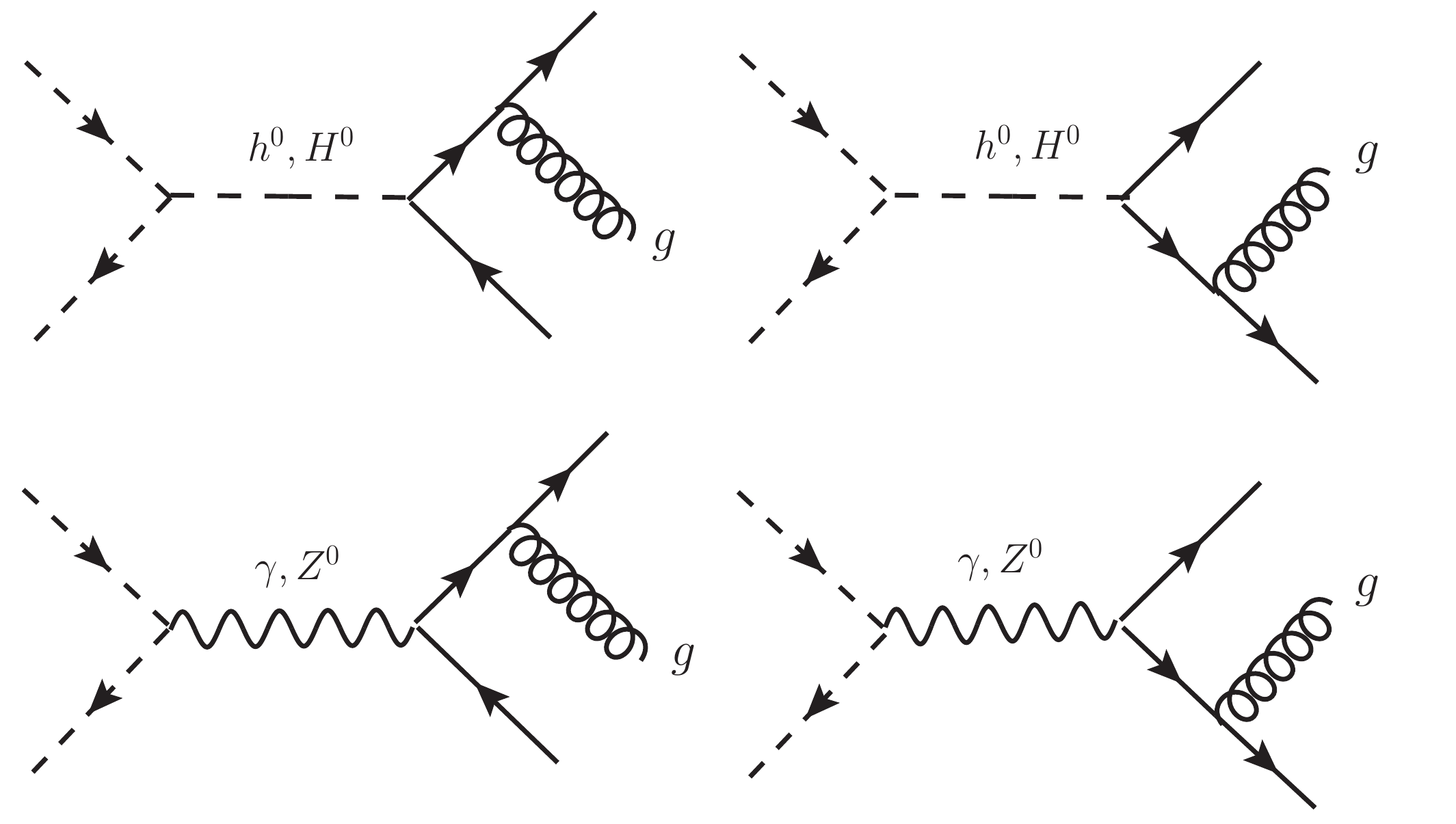}
    \caption{Real gluon emission to the processes of Fig.\ \ref{fig:tree}.}
    \label{fig:real}
\end{figure}

The $\mathcal{O}(\alpha_s)$ contributions are completed by the gluon radiation diagrams shown in Fig.\ \ref{fig:real}, which cancel the infrared divergences introduced by the virtual corrections that include the exchange of a massless gluon. In order to combine the virtual and the real corrections, cancelling the infrared divergences, we make use of the dipole subtraction method \cite{Catani1996, Catani2002}. This method is based on the construction of an auxiliary cross section $\sigma_{\rm A}$, which includes the information about the infrared divergent behaviour of the original cross section. Moreover, the auxiliary cross section is constructed such that the one-particle gluon phase space is factorized from the three particle phase-space and can be integrated out analytically. Using the auxiliary cross section, the next-to-leading order (NLO) cross section can be written as
\begin{align}  
    \begin{split}
    \sigma_{\rm NLO} ~=~& \sigma_{\rm LO} ~+~ 
        \int_{2 \to 3} \Bigr[ {\rm d}\sigma^{\rm R}_{\epsilon=0} 
            - {\rm d}\sigma^{\rm A}_{\epsilon=0} \Bigr] \\
            &~+~ \int_{2 \to 2} \biggr[ {\rm d}\sigma^{\rm V} + \int_1 {\rm d}\sigma^{\rm A} \biggr]_{\epsilon=0} \,,
    \label{Eq:sigNLO}
    \end{split}
\end{align}
where $\sigma_{\rm LO}$ is the leading-order cross section, ${\rm d}\sigma^{\rm R}$ and ${\rm d}\sigma^{\rm V}$ are the differential cross sections stemming from the real emission and vertex correction diagrams, respectively. The first integration in Eq.\ \eqref{Eq:sigNLO} is performed over the three-particle phase space corresponding to the real emission diagrams, and the second integration is performed over the two-particle phase space. For this last part, the auxiliary cross section is integrated analytically over the gluon phase space. For the explicit construction of the auxiliary cross section we refer the reader to our previous work \cite{ChiChi2qq3}.

As mentioned above, the ultraviolet divergences are removed by a proper redefinition of model parameters which requires a careful definition of these parameters, i.e., choosing a renormalization scheme. In our previous works \cite{ChiChi2qq3, NeuQ2qx1, NeuQ2qx2} we have proposed and used a mixed on-shell and \DRbar\ renormalization scheme of the MSSM which is well-suited for dark matter calculations. However, there is a certain ambiguity in choosing the renormalization scheme, which we want to demonstrate here by defining an alternative scheme. The new alternative renormalization scheme differs from our standard scheme just by having the top quark mass defined in the \DRbar\ scheme. The alternative renormalization scheme is also particularly suitable to study variations of the renormalization scale as all relevant parameters are defined in the \DRbar\ scheme. The differences resulting from using the two alternative schemes or from varying the renormalization scale will be discussed in Sec.~\ref{Sec:RelicIa}.

\subsection{QED Sommerfeld corrections}
\label{Sec:Sommerfeld}

\begin{figure}
 \includegraphics[width=0.49\textwidth]{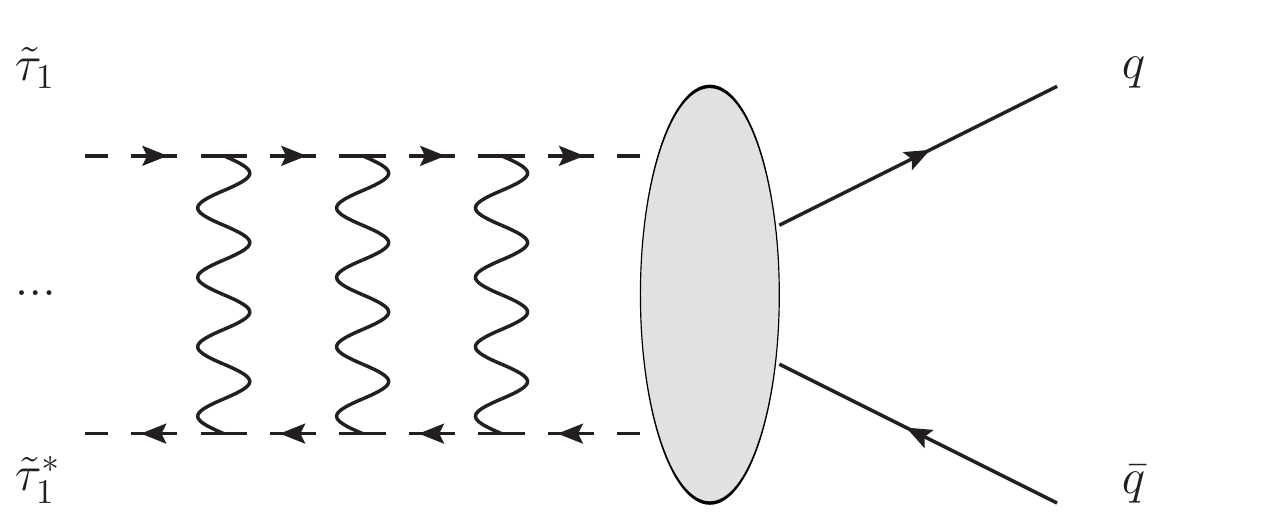}
 \caption{Multiple photon exchange in the initial state leading to the Sommerfeld enhancement.}
 \label{Fig:Sommerfeld}
\end{figure}

In the limit of low relative velocities, as it is typical during freeze-out of dark matter, annihilating particles can exchange light mediators leading to the well-known Sommerfeld enhancement. In our case this effect is caused by the exchange of multiple photons between the incoming stau anti-stau pair, see Fig.~\ref{Fig:Sommerfeld}. For an each exchange of a photon the cross section is corrected by a factor proportional to $(\alpha/v_{\mathrm{rel}})$. With $\alpha \approx v_{\mathrm{rel}}$ during freeze-out, this contribution becomes non-perturbative and thus has to be resummed to all orders of perturbation theory. 

The Sommerfeld effect in electroweak theories has been discussed intensively in the literature \cite{Slatyer:2009vg, Drees:2009gt, Feng:2010zp, Hryczuk:2011vi, Hryczuk:2011tq, Beneke2014a, Beneke2014b} and was studied previously for QCD in the context of DM@NLO to which we refer for more details regarding the computation and implementation \cite{StSt2xx,StSt2qq,scalevar}.

As stau annihilation occurs only via an $s$-channel exchange, the $s$-wave contribution dominates the squared matrix element. Therefore, we can factorize the corrected cross section in terms of the leading order contribution
\begin{align}
\label{eq:sumfactor}
    \left(\sigma v\right)_{\mathrm{resum}} = S_{0} \left(\sigma v\right)_{\mathrm{tree}}\,,
\end{align}
with the Sommerfeld factor $S_{0}$. The latter is evaluated by solving the Schr\"odinger equation 
\begin{align}
\label{eq:SchroedingerEQ}
	\Big[ -\frac{4}{m_{\tilde{\tau}_1}} \nabla^2 + V(\mathbf{r}) - \big( E+i\Gamma_{\tilde{\tau}_1} \big) \Big]
		&\mathcal{G} \big( \mathbf{r};E +i\Gamma_{\tilde{\tau}_1}\big)\nonumber \\
		&= \delta^{(3)}(\mathbf{r})\,,
\end{align}
with $E=\sqrt{s}-2m_{\tilde{\tau}_1}$ and the Coulomb potential \cite{Billoire:1979ih, Kauth:2011bz}
\begin{align}
	V(\mathbf{r}) =& \,-\frac{\alpha(\mu_C)}{\mathbf{r}} \\  &\times \bigg\{ 1 + \frac{\alpha(\mu_{C})}{4\pi} \bigg[2 \beta_0 \bigg(\ln(\mu_C \mathbf{r}) + \gamma_E \bigg) + a_1 \bigg] \bigg\}\,. \nonumber
    \label{eq:Coulombpotential}
\end{align}
The solution is the Green's function $\mathcal{G}(\mathbf{r};E+i\Gamma_{\tilde{\tau}_1}) = \mathcal{G}(\mathbf{r}, \mathbf{r}^\prime=0;E+i\Gamma_{\tilde{\tau}_1})$. $\gamma_E = 0.5772$ indicates the Euler-Mascheroni constant,
\begin{align}
	a_1 &= -\frac{20}{9} \sum_{f} Q_f^2\,, 
\end{align}
and
\begin{align}
	\beta_0 &= -\frac{4}{3} \sum_{f} Q_f^2\,, 
\end{align}
are parameters whereas the latter originates from the one-loop $\beta$-function including all fermions $f$ up to the scale of the typical momentum exchange. The typical scale is taken to be the Coulomb scale $\mu_C$,
\begin{align} 
    \mu_C = \mathrm{max} \left\{\mu_B , 2m_{\tilde \tau_1} v_s \right\} \,,
\end{align}
calculated as the maximum value of the momentum exchange of the unbound particles and the Bohr momentum, $\mu_B = 2m_{\tilde \tau_1}\alpha$. The velocity $v_s$ indicates the non-relativistic velocity of one of the incoming staus ($v_{\mathrm{rel}}=2v_s$). Finally, the Sommerfeld factor in Eq.~\eqref{eq:sumfactor}, which multiplies the tree-level cross section, can be evaluated as the ratio of the two Green's functions at the origin ($\bf r = 0$) \cite{Cassel:2009wt,Hagiwara:2008df} 
\begin{align}
    \label{eq:Sommerfeldfactor}
	S_{0} = \frac{\Im \left[\mathcal{G}\left(\mathbf{0};E+i\Gamma_{\tilde{\tau}_1}\right)\right]}{\Im\left[\mathcal{G}_0\left(\mathbf{0};E+i\Gamma_{\tilde{\tau}_1}\right)\right]} \,,
\end{align}
where the Green's function $\mathcal{G}_0(\mathbf{0},E+i\Gamma_{\tilde{t}_1})$ stands for the solution of the Schr\"odinger equation without any Coulomb potential,
\begin{align}
    \Im \left[\mathcal{G}_0\left(\mathbf{0};E+i\Gamma_{\tilde{\tau}}\right)\right] = \frac{m^2_{\tilde{\tau}_1} v_{\rm s}}{4 \pi}\,.
\end{align}
Given the negligible scale dependence of $\alpha_{\mathrm{em}}$, its running can be neglected for the calculation of the Sommerfeld factor. As the Coulomb potential given in Eq.\ \eqref{eq:Coulombpotential} is scale independent by itself, this implies similarly a negligible contribution of the $\beta$-function. With the NLO contribution being suppressed by an additional factor of $\alpha/(4 \pi)$, it has generally only a negligible effect on the correction. Hence, we performed our final calculation by including the Coulomb potential at leading order only. For further details on the numerical evaluation, we refer to our previous papers \cite{StSt2xx, StSt2qq, scalevar}. 

\begin{figure*}[t]
    \centering
    \includegraphics[width=0.49\textwidth, trim={0.3cm 0.3cm 0 0}, clip]{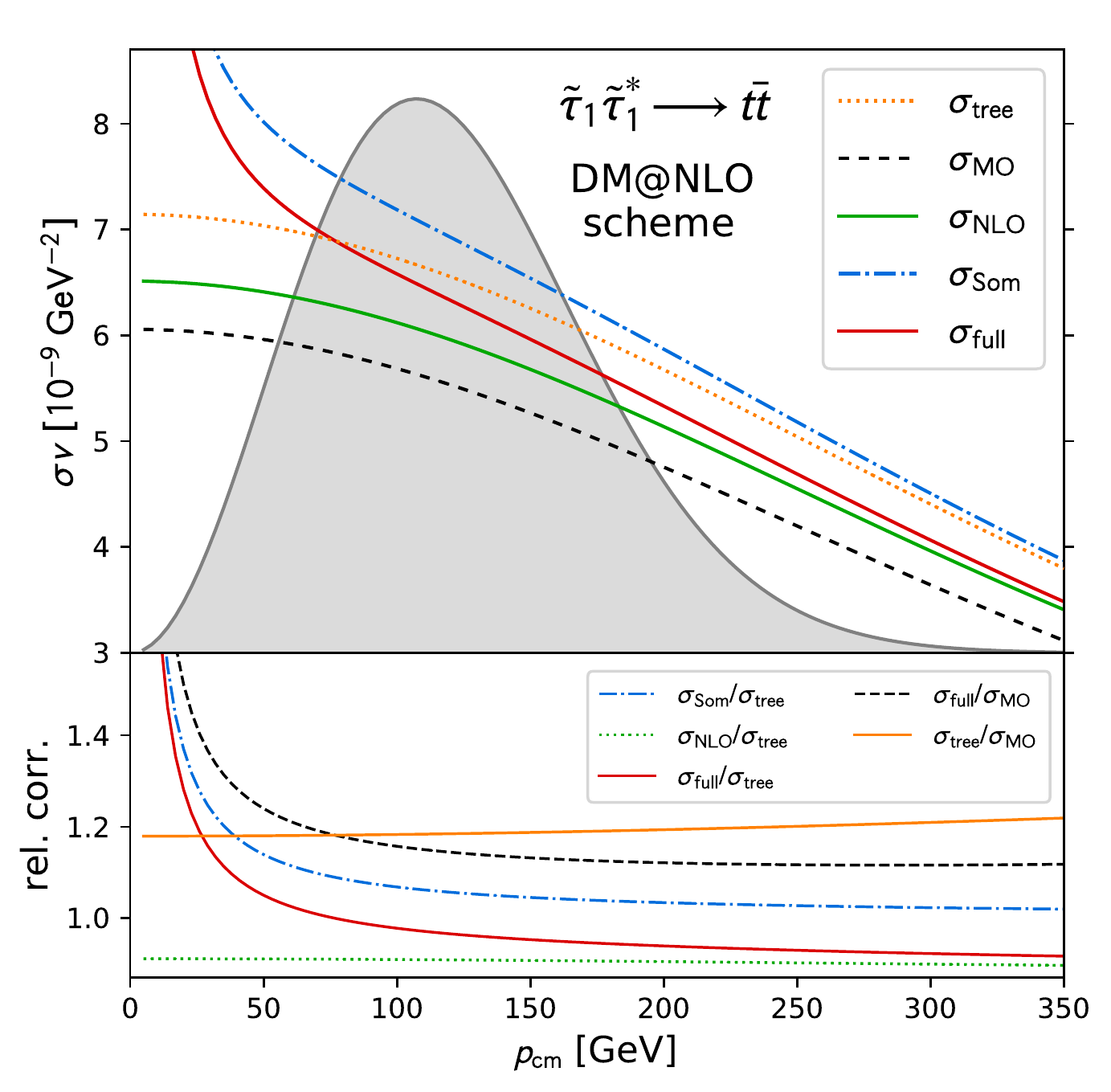}~~~
    \includegraphics[width=0.49\textwidth, trim={0.3cm 0.3cm 0 0}, clip]{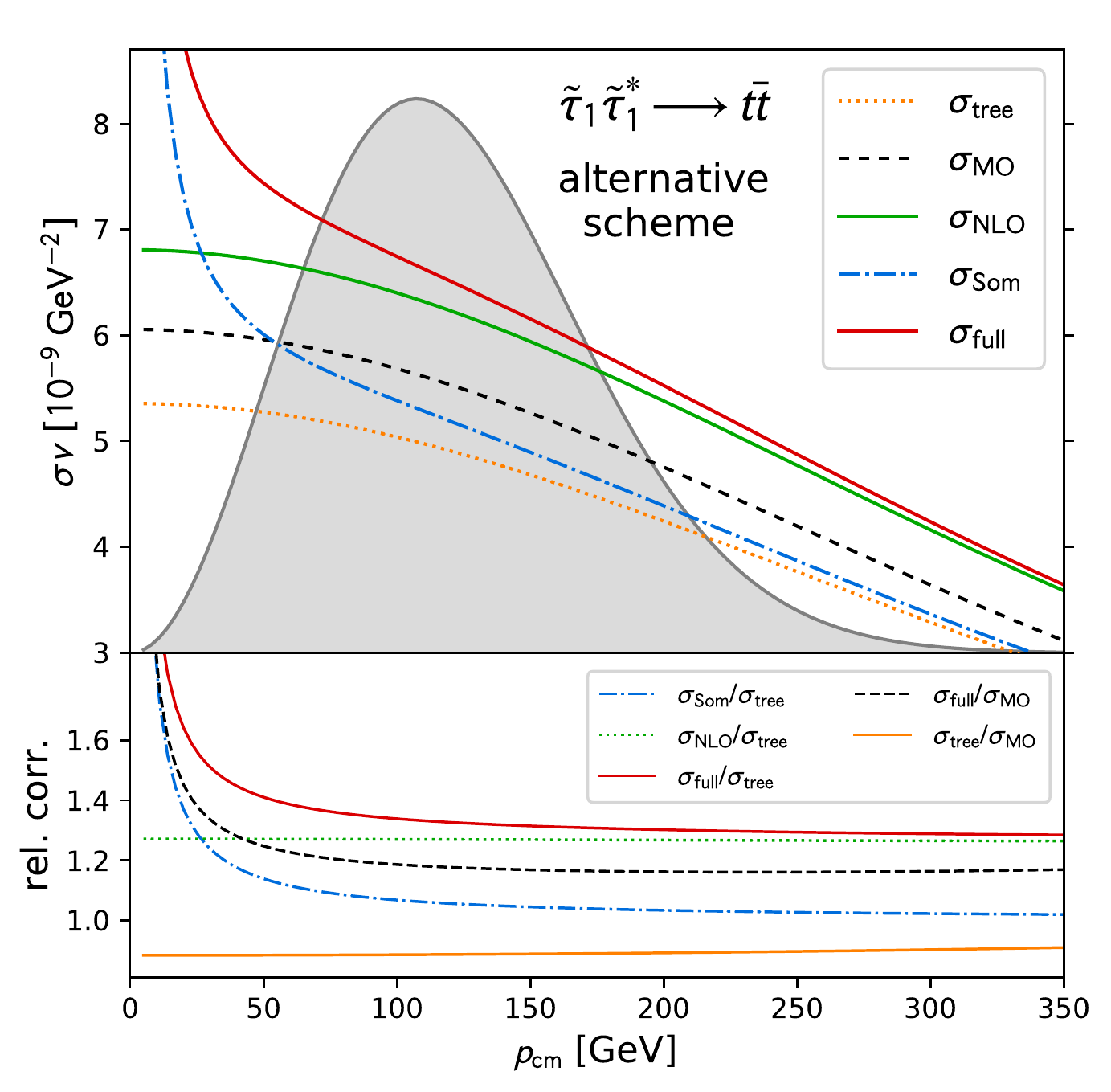}
    \caption{Annihilation cross section of the process $\tilde{\tau}_1 \tilde{\tau}_1^* \to t\bar{t}$ as a function of the center-of-mass momentum $p_{\rm cm}$ for Scenario I of Tab.\ \ref{Tab:Scenarios} using the standard \DMNLO\ renormalization scheme (left) and the alternative scheme (right). The upper panels show tree-level results and different levels of corrections as discussed in Sec.\ \ref{Sec:Calculation}. The lower panels show the corresponding relative corrections. The grey areas indicate the thermal distribution in arbitrary units.}
    \label{Fig:XSecIa}
\end{figure*}
\section{Numerical results}
\label{Sec:Results}

Let us now discuss the numerical impact of the corrections presented in Secs.\ \ref{Sec:NLO} and \ref{Sec:Sommerfeld}, first on the annihilation cross section itself, and then on the prediction for the relic density of dark matter. For this numerical study, we will rely on the two reference scenarios defined in Tab.\ \ref{Tab:Scenarios} and discussed in Sec.\ \ref{Sec:Pheno}.

In order to compute the relic density including the corrections discussed above, our full NLO calculation including Sommerfeld corrections has been implemented in the \DMNLO\ package. In practice, the evaluation of the Boltzmann equation by {\tt micrOMEGAs} uses cross sections computed by {\tt CalcHEP} which are replaced in specific cases by the values obtained from the {\tt DM@NLO} calculation. In this way, the processes included in the {\tt DM@NLO} are taken into account in a consistent way throughout the calculation of the relic density and provide a more precise prediction of the relic density.

\begin{figure*}
    \centering
    \includegraphics[width=0.49\textwidth, trim={0 0.3cm 0 0}, clip]{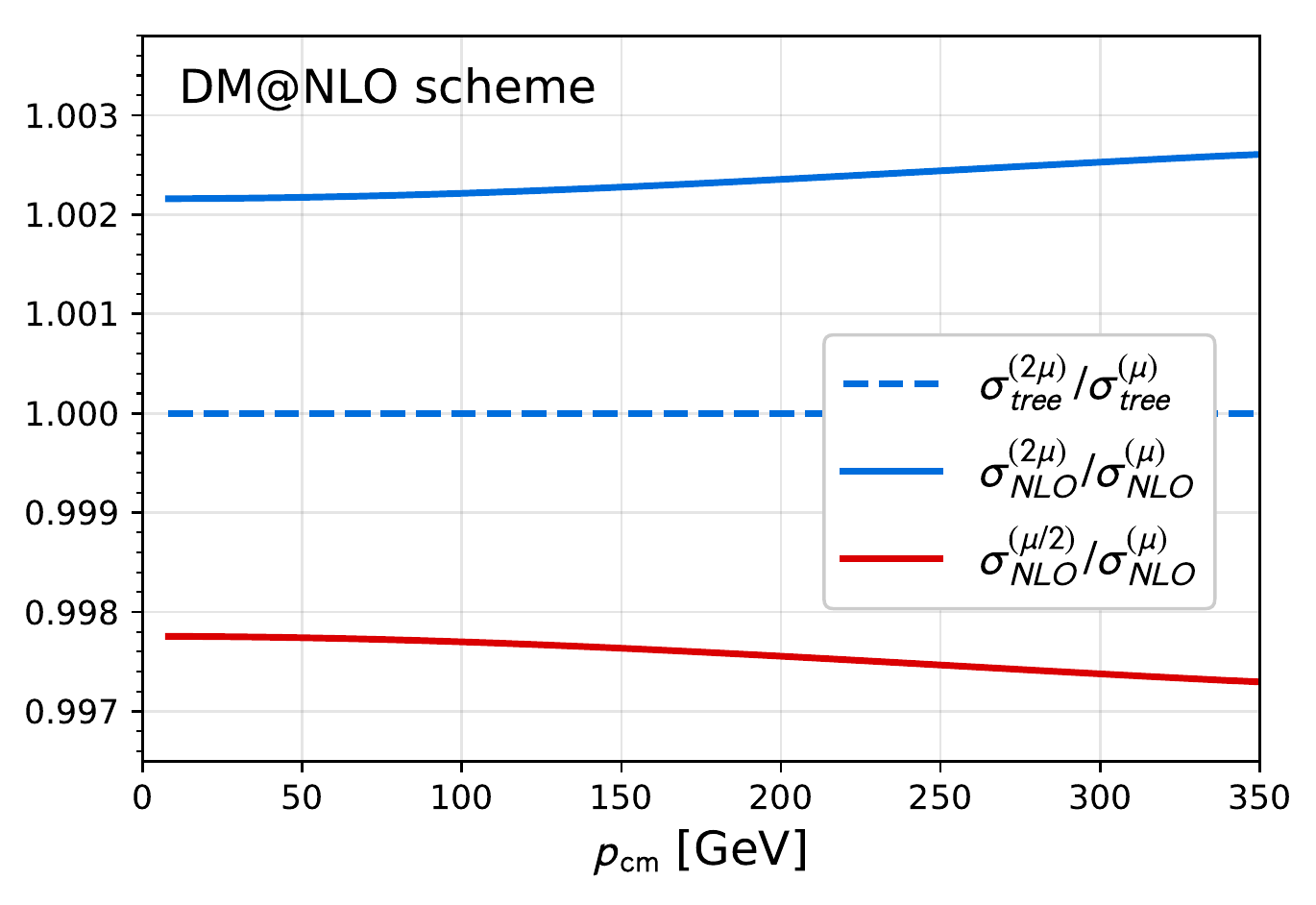}~~~
    \includegraphics[width=0.4825\textwidth, trim={0 0.3cm 0 0}, clip]{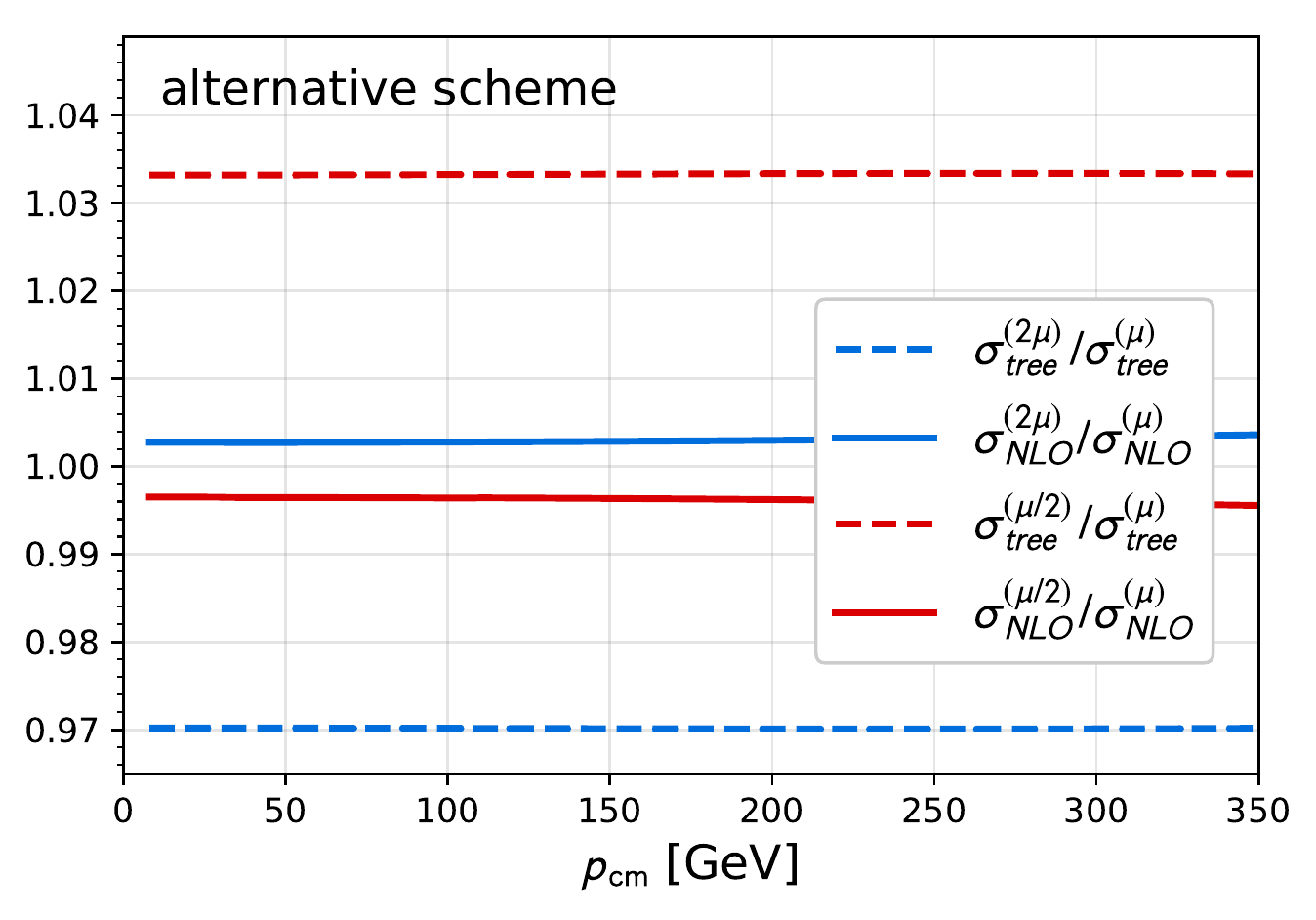}
    \caption{Ratios of the $\tilde\tau_1 \tilde \tau_1^{*} \rightarrow t \bar t$ cross section for renormalization scale $\mu$ varied around the central scale ($\mu=$1 TeV) at leading (dashed line) and next-to-leading order (solid line) for the \DMNLO\ renormalization scheme (left) and the alternative scheme (right).}
    \label{Fig:scale}
\end{figure*}

\subsection{Annihilation cross section and its theoretical uncertainty}
\label{Sec:XSec}

\begin{figure}
    \centering
    \includegraphics[width=0.49\textwidth, trim={0 0.3cm 0 0}, clip]{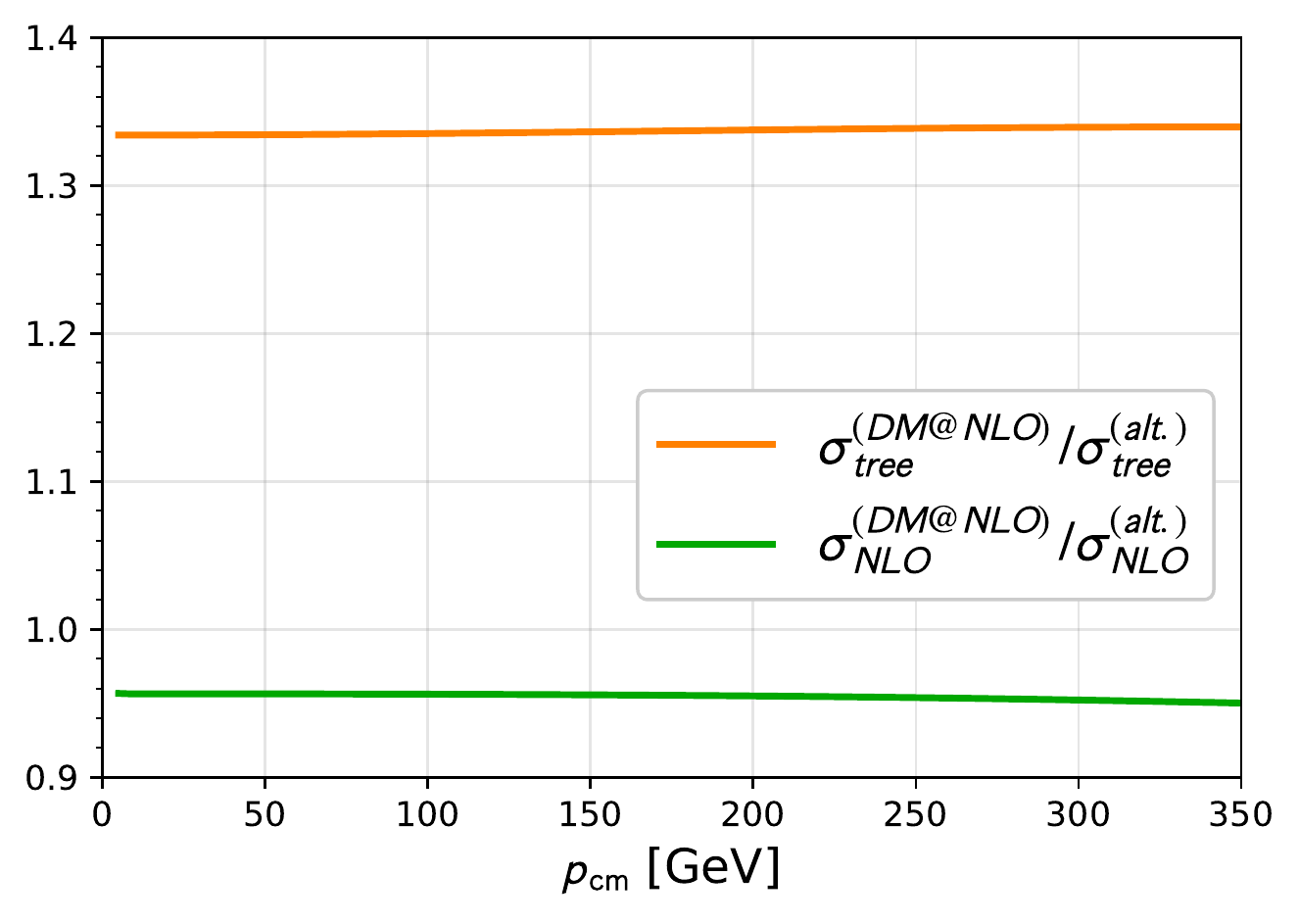}
    \vspace*{-2mm}
    \caption{Ratios of the $\tilde\tau_1 \tilde \tau_1^{*} \rightarrow t \bar t$ cross section calculated in the \DMNLO\ and the alternative renormalization schemes at leading (orange) and at next-to-leading-order (green).}
    \label{Fig:scheme}
\end{figure}

In Fig.\ \ref{Fig:XSecIa} we show the stau-antistau annihilation cross section as a function of the center-of-mass momentum $p_{\rm cm}$ for masses and couplings from the Scenario I of Tab.\ \ref{Tab:Scenarios}. Given that in the Boltzmann equation the total cross section is thermally averaged, we also show the corresponding thermal distribution. The velocity distribution indicates the momentum range which is most relevant for the computation of the relic density.

The two different plots in Fig.\ \ref{Fig:XSecIa} show the next-to-leading-order annihilation cross section results for both renormalization schemes mentioned in Sec.~\ref{Sec:NLO}. Let us first discuss the results using our standard \DMNLO\ renormalization scheme. We compare our results to the result from {\tt micrOMEGAs} (black dashed line). We see that our leading-order (LO) result (orange dotted line) does not coincide with the {\tt micrOMEGAs} cross section. One of the reasons is the different definition of the top quark mass. In the \DMNLO renormalization scheme we use the physical on-shell top quark mass whereas {\tt micrOMEGAs} uses the top quark mass in the \DRbar\ scheme. The other reason is the difference in the Yukawa couplings due to the fact that {\tt micrOMEGAs} uses effective couplings to include some higher-order corrections. 

Including the NLO corrections decreases the cross section by about 9\% as compared to the LO result, while the NLO cross section is about 7.4\% larger compared to the {\tt micrOMEGAs} result. The relative correction is fairly constant for a large span of the center-of-mass momentum $p_{\rm cm}$. On top of the next-to-leading-order SUSY-QCD corrections, we include also the electroweak Sommerfeld enhancement.

The Sommerfeld enhancement dominates the cross section for small relative velocities. For an attractive force such as the electromagnetic force between a stau and an anti-stau particle, the Sommerfeld enhancement increases the cross section (blue dash-dotted line in Fig.\ \ref{Fig:XSecIa}). The final correction to the leading-order cross section (red line in Fig.\ \ref{Fig:XSecIa}) after including both the SUSY-QCD NLO corrections and the electroweak Sommerfeld enhancement is relatively small given that both effects compensate each other.

In addition to the shift of the numerical result, including higher-order corrections leads to a better estimate of the theoretical uncertainty associated with the prediction. The theoretical uncertainty is the estimate of the contributions of higher orders that are not included in the actual calculation. There are several methods to estimate this uncertainty.

One possibility relies on the fact that the dependence on the renormalization scale introduced through the higher-order corrections would disappear if all orders in perturbation theory could be included. The dependence on the renormalization scale is gradually reduced by including higher-order corrections. That means in turn that the remaining dependence is an estimator for the scale-dependent parts of the missing higher-order contributions. 

We have investigated the dependence of the cross section on the variation of the renormalization scale for both renormalization schemes (for technical details see Ref.\ \cite{scalevar}). The results are shown in Fig.\ \ref{Fig:scale}. In the left panel, we show the impact of the variation of the renormalization scale between $\mu = 0.5$ TeV and $\mu = 2$ TeV on the next-to-leading-order cross section calculated in the \DMNLO\ renormalization scheme. The leading-order cross section is completely insensitive to the scale variation and even the next-to-leading-order cross section is only mildly sensitive in this scheme. This is simple to understand as the most prominent parameter in this case, the top quark mass, is defined in the on-shell scheme which by definition removes the renormalization scale dependence related to the top quark mass from both the leading and next-to-leading-order cross sections. The dependence of the next-to-leading-order cross section on the renormalization scale comes from the scale dependent strong coupling constant which was first introduced by the SUSY-QCD one-loop corrections. But even this dependence is only mild due to the high scale of $\mu = 1$ TeV which is natural for this process. At such high scales, the changes in $\alpha_s$ due to the change in scale are very small.

\begin{figure}
    \centering
    \includegraphics[width=0.5\textwidth, trim={0.2cm 0.05cm 0 0.8cm}, clip]{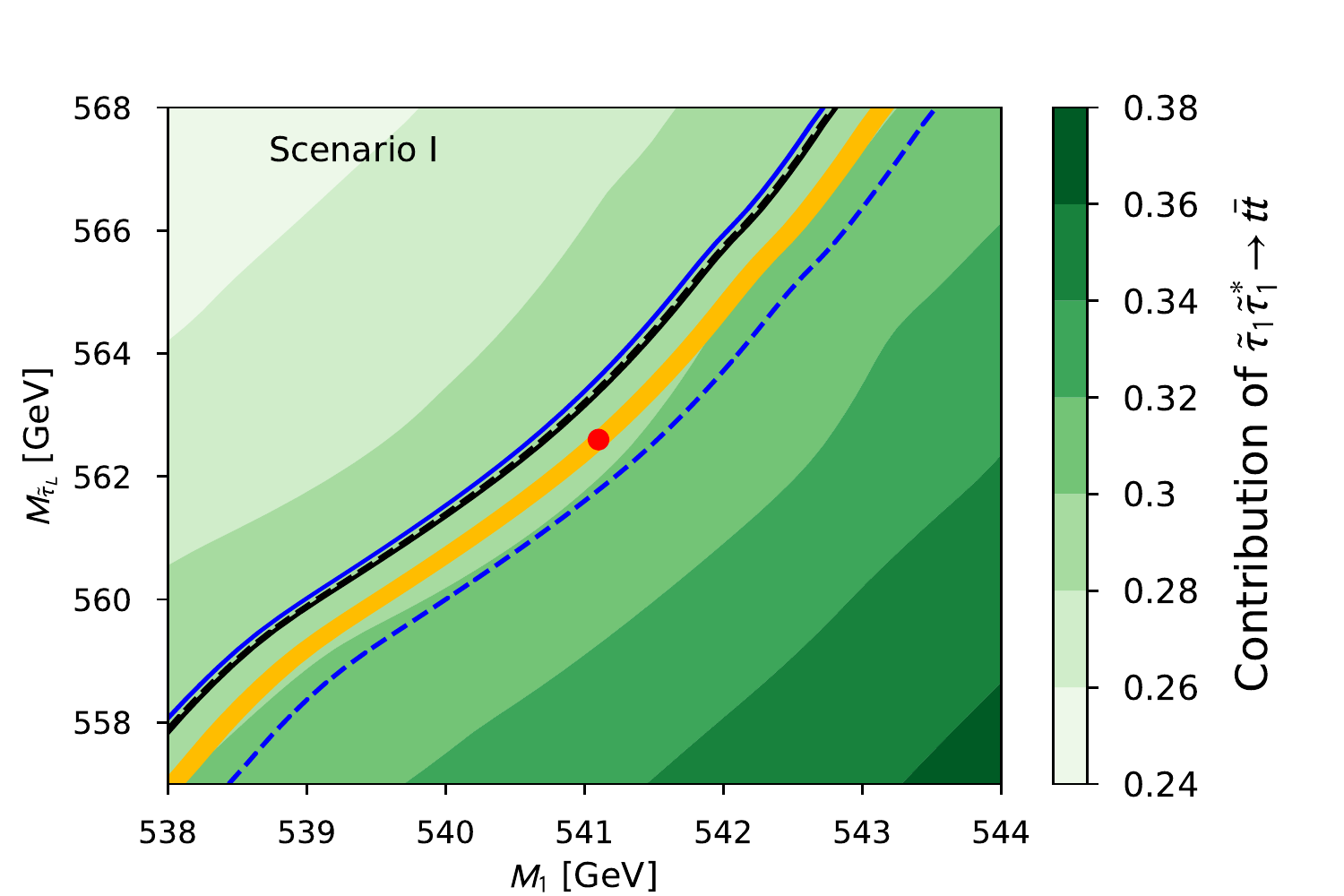}
    \vspace*{-2mm}
    \caption{Comparison of experimental and theoretical uncertainties in the $M_1$--$M_{\tilde{\tau}_L}$ plane around reference Scenario I (indicated by the red dot). The yellow band shows the experimental uncertainties given in see Eq.\ \eqref{Eq:Planck} as measured by the Planck satellite at the 1$\sigma$ confidence level. The leading (next-to-leading) order relic density from both our renormalization schemes is denotes by blue (black) lines. The predictions in the \DMNLO\ (alternative) renormalization scheme are shown using the solid (dashed) lines. As in Fig.\ \ref{Fig:Omh2}, the green contours indicate the relative contribution of the process $\tilde{\tau}_1\tilde{\tau}_1^* \to t\bar{t}$ to the total annihilation cross section, based on the {\tt micrOMEGAs} calculation.}
    \label{Fig:Omh2theouncert}
\end{figure}

In the case of the NLO calculation in the alternative renormalization scheme, the top quark mass was defined in the \DRbar\ scheme which leads to larger sensitivity to the change in the renormalization scale. As one can see in the right panel of Fig.\ \ref{Fig:scale}, the scale dependence in the alternative scheme is about 3\% at leading order and is reduced to per mille level at NLO.

\begin{figure*}
    \centering
    \includegraphics[height=0.37\textwidth, trim={0.2cm 0 3.5cm 0}, clip]{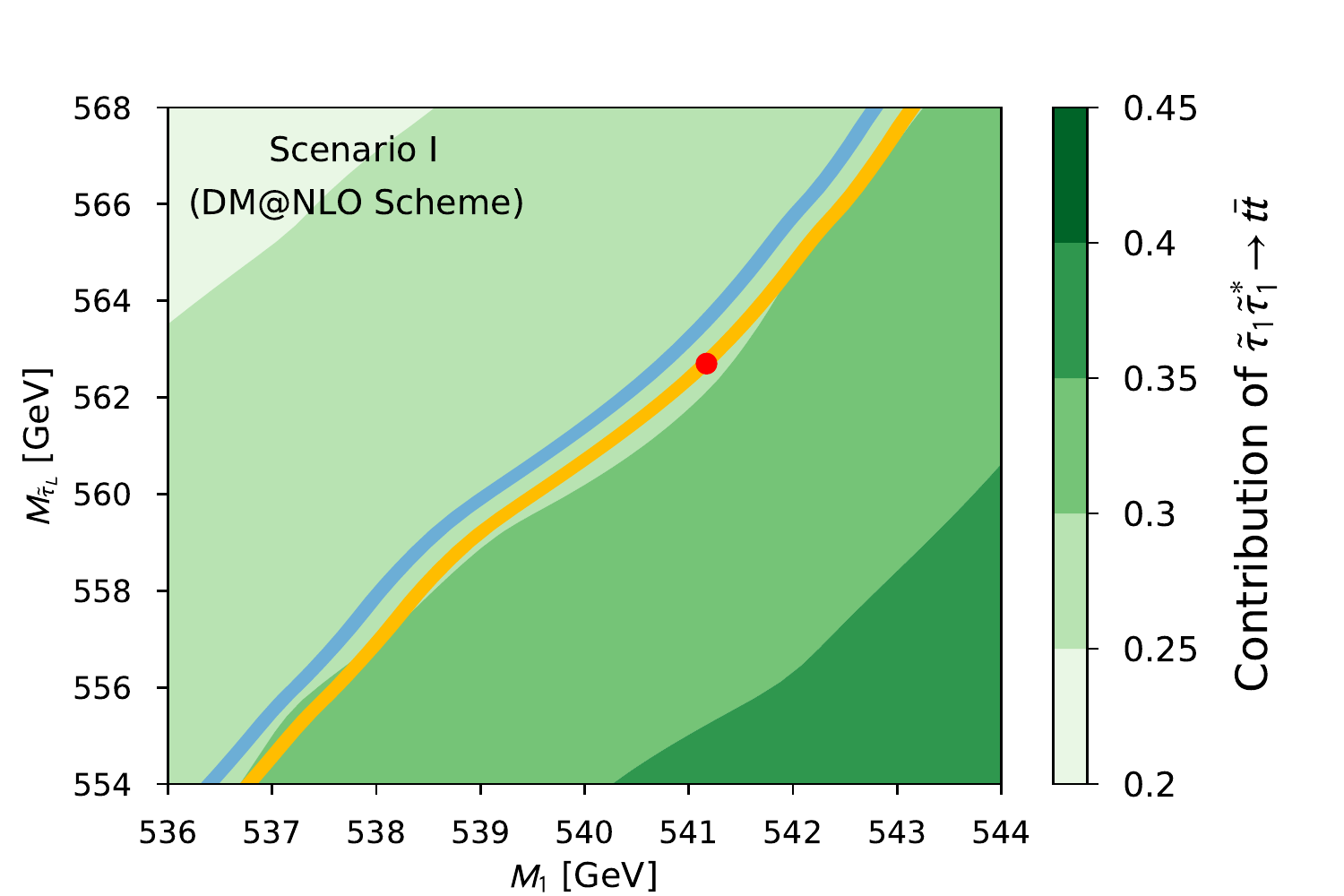}~~~
    \includegraphics[height=0.37\textwidth, trim={0.2cm 0 0.8cm 0}, clip]{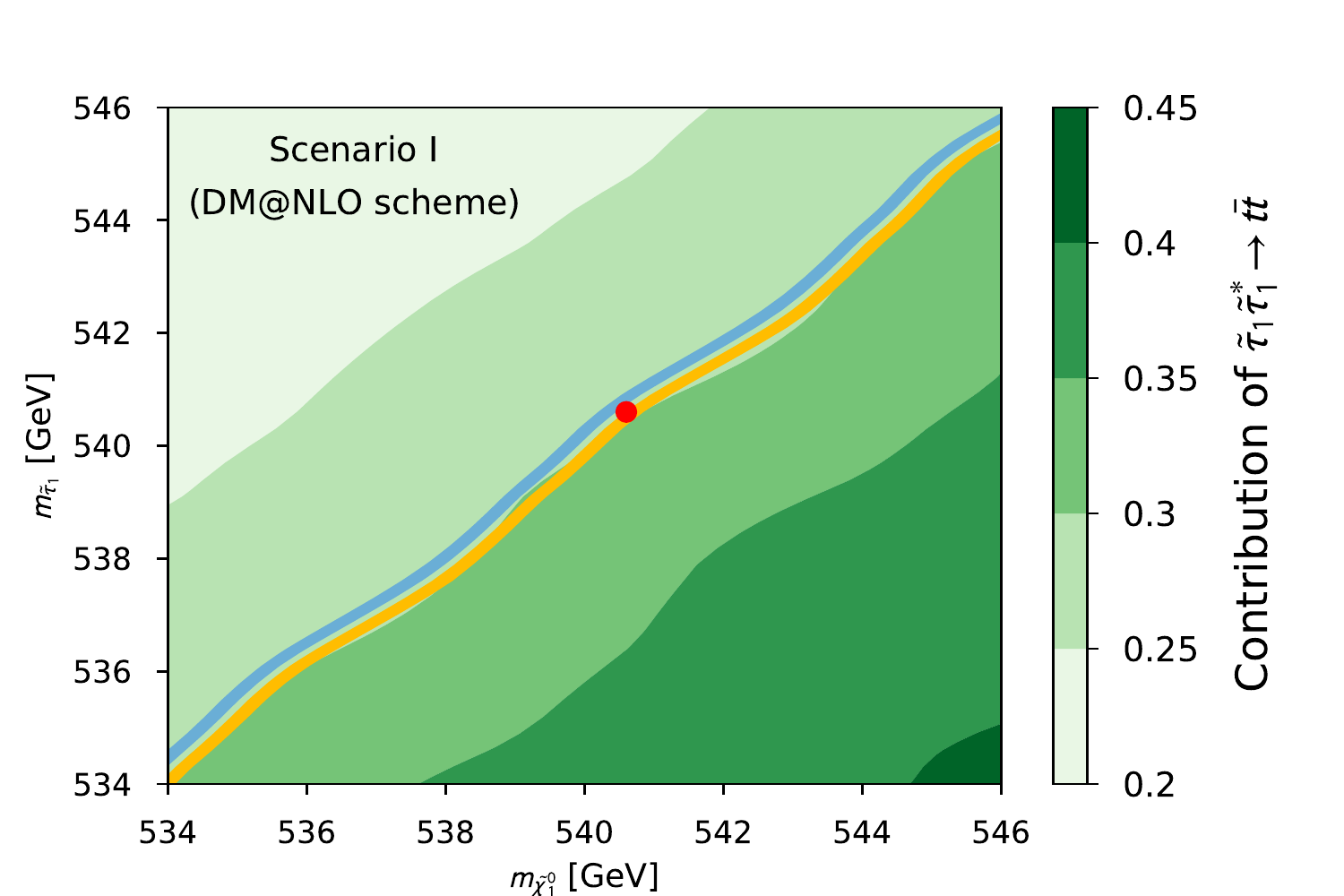}
    \vspace*{-2mm}
    \caption{Parameter regions in the $M_1$--$M_{\tilde{\tau}_L}$ plane (left) and  $m_{\tilde{\chi}^0_1}$--$m_{\tilde{\tau}_1}$ plane (right) that are compatible with the Planck limits given in Eq.\ \eqref{Eq:Planck}, where the stau relic density has been computed using {\tt micrOMEGAs} (orange) and our full NLO and Sommerfeld corrected cross section (blue). All other parameters are fixed to those given for Scenario I in Tab.\ \ref{Tab:Scenarios}. The red dot correspond to the Scenario I. The green contours correspond to the relative contribution of the process $\tilde{\tau}_1 \tilde{\tau}_1^* \to t\bar{t}$ to the total annihilation cross section.}
    \label{Fig:Omh2NeuIa}
\end{figure*}

From the investigation of the dependence of the cross section on the renormalization scale, we might conclude that the theoretical uncertainty at NLO is smaller than three per mille. There are some caveats to this conclusion. First, as we have seen, the sensitivity to scale changes depends on the renormalization scheme where only pure \MSbar\ or \DRbar\ schemes exhibit the full sensitivity. The other important caveat is that varying the renormalization scale highlights only the size of the scale-dependent part of the higher-order corrections. In order to highlight the shortcomings of the estimation of the theoretical uncertainty by varying the renormalization scale, we compare the changes in the cross sections due to different renormalization schemes. A renormalization scheme is specified by the definition of the model parameters and the corresponding definition of the model parameter counterterms. At leading order the different definition of parameters cause a large difference between calculations in different renormalization schemes. This can be seen either by comparing the left and right panels in Fig.\ \ref{Fig:XSecIa} or by constructing the ratio of the leading-order cross sections as in Fig.\ \ref{Fig:scheme}. In our case the difference between the leading-order cross sections is larger than 30\%. At next-to-leading order the counterterms compensate for the difference in parameter definitions. The only difference between the next-to-leading order calculations in different renormalization schemes comes from the use of different parameters in the one-loop corrections. This difference is of a higher order and can be used as an estimate of theoretical uncertainty. In our case the difference between the NLO predictions in our two schemes is only about 4-5\%. The theoretical uncertainty defined in this way also reduces with every order included in the calculation and it takes into account not only terms sensitive to changes in the renormalization scale but all terms which depend on the model parameters e.g. the masses. This definition of theoretical uncertainties is also not without flaws. To truly assess the theoretical uncertainty one would ideally use many different renormalization schemes which are not always simple to define consistently for a given model (here the MSSM). And even if this was feasible, this approach as well as the previous one, cannot capture the presence of constant terms which can be determined only by an exact calculation of the higher-order corrections whose size we are trying to estimate.

We see that in order to be conservative, in our case we should choose the variation of the renormalization scheme to define the theoretical uncertainty. We then conclude that the leading-order cross section has an uncertainty of about 30\% and the cross section including the next-to-leading order corrections has still an uncertainty of 4-5\%. Using the \DMNLO\ renormalization scheme produces smaller higher-order corrections indicating quicker convergence of the perturbative series. This is one of the reasons why we adopt this scheme again in the following and we apply it to the relic density calculation, assuming neutralino or gravitino dark matter within the pMSSM.

\subsection{Impact on the neutralino relic density}
\label{Sec:RelicIa}

We first consider the case, where the lightest neutralino is the dark matter candidate and the second-lightest supersymmetric particle is the lighter stau. This situation corresponds to the mass spectrum of Scenario I defined in Tab.\ \ref{Tab:Scenarios}.

In order to study the impact of the higher-order corrections on the relic density, we vary two key parameters, namely the bino mass parameter $M_1$ and the left-handed stau mass parameter $M_{\tilde{\tau}_L}$ around the values specified in the Scenario I. The parameter region where the relic density satisfies the experimental constraint given by Eq.\ \eqref{Eq:Planck} is shown in Fig.\ \ref{Fig:Omh2theouncert} as a yellow band. The band is determined using the {\tt micrOMEGAs} relic density calculation of the cross section. The width of the band corresponds to one sigma experimental uncertainty. The blue solid and dashed lines in Fig.\ \ref{Fig:Omh2theouncert} denote the predictions for the correct relic density from the leading-order calculations in the \DMNLO\ and the alternative renormalization scheme, respectively. The band formed by these two lines denotes the theoretical uncertainty of the leading-order relic density calculation. Similarly, the black solid and dashed lines correspond to the next-to-leading-order relic density prediction in the two schemes. We see that the NLO predictions from both renormalization schemes are very consistent with each other and the theoretical uncertainty of the relic density determination at NLO is very small in this scenario.

In Fig.\ \ref{Fig:Omh2NeuIa} we first compare the final relic density prediction from {\tt micrOMEGAs} and from the NLO calculation in the \DMNLO\ scheme in the plane of the soft mass parameters $M_1$ and $M_{\tilde{\tau}_L}$, then in the plane of the corresponding physical neutralino and stau masses. Note that, also for the right plot of Fig.\ \ref{Fig:Omh2NeuIa}, all other input parameters are fixed to the values given in Tab.\ \ref{Tab:Scenarios}.

We observe that the shift between the favoured region based on the {\tt micrOMEGAs}/{\tt CalcHEP} calculation and the one based on our full calculation amounts to a shift of about 3 GeV for the neutralino mass (for fixed stau mass) or about 5 GeV for the stau mass (for fixed neutralino mass). Most importantly, the shift is much larger than the width of the respective band corresponding to the experimental uncertainty given in Eq.\ \eqref{Eq:Planck}. This shows the importance of including higher-order corrections, and in particular the importance of including the Sommerfeld enhancement in the present situation.

\subsection{Impact on the gravitino relic density}
\label{Sec:RelicIb}

Here, we consider the case of the gravitino being the lightest supersymmetric particle. The second-lightest particle is the lighter stau $\tilde{\tau}_1$ with the mass given in Tab.\ \ref{Tab:Scenarios} for Scenario II. Let us recall that in this illustrative scenario, all other superpartners are rather heavy with masses of about 5 TeV to simplify the analysis. 
 
\begin{figure}
    \centering
    \includegraphics[width=0.5\textwidth, trim={0.3cm 0.3cm 0 1.2cm}, clip]{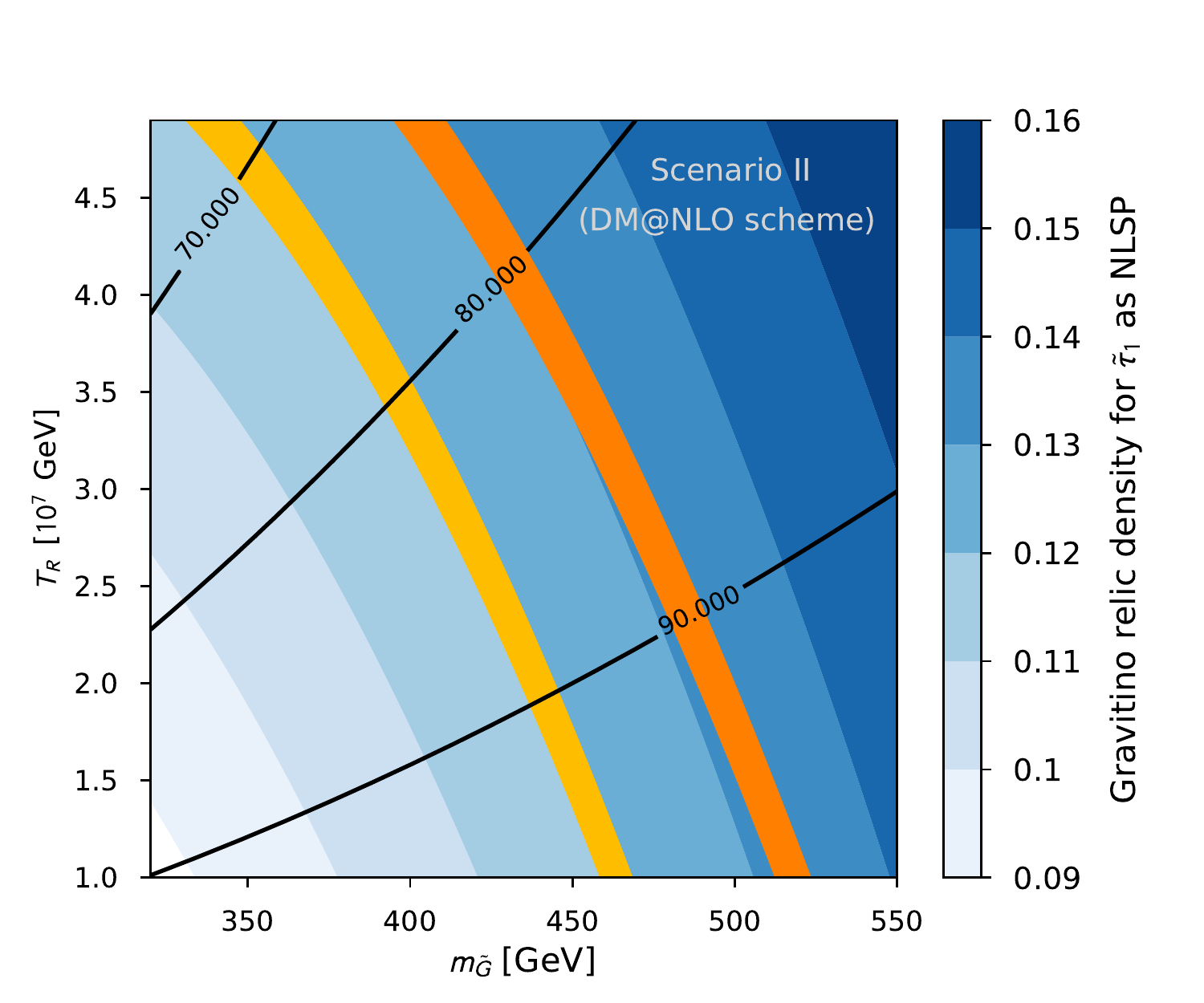}
    \caption{Parameter regions in the $m_{\tilde{G}}$--$T_R$ plane which are compatible with the Planck limits given in Eq.\ \eqref{Eq:Planck} for the case of gravitino dark matter, where the stau relic density has been computed using {\tt micrOMEGAs} (yellow) and our full NLO and Sommerfeld corrected cross section in the {\tt DM@NLO} scheme (orange). All other parameters are fixed to those given for Scenario II in Tab.\ \ref{Tab:Scenarios}. The blue contours correspond to the gravitino relic density based on the {\tt micrOMEGAs} calculation. The black lines indicate the relative non-thermal contribution in percent according to Eq.\ \eqref{Eq:Omh2GravNonTh} to the total gravitino relic density, again based on the {\tt micrOMEGAs} calculation.}
    \label{Fig:Omh2Grav}
\end{figure}

In a similar way as above for neutralino dark matter, we illustrate in Fig.\ \ref{Fig:Omh2Grav} the impact of our NLO and Sommerfeld corrections presented in Sec.\ \ref{Sec:NLO} on the favoured region of parameter space. As in the previous case, the shift between the {\tt micrOMEGAs} calculation and our full calculation is more important than the Planck uncertainty given in Eq.\ \eqref{Eq:Planck}. 

Although the non-thermal contribution accounts for only about 80\% of the gravitino relic density, and the process affected by the presented corrections accounts for only about 32\% of the total stau annihilation cross-section, the observed shift is more important than the impact found for Scenario I. This is caused by a relatively large impact of the NLO corrections in this scenario. Here, contrary to Scenario I, the squarks and gluino are rather heavy, such that the corresponding gluino loop contribution (see Fig.\ \ref{fig:virtual}) is suppressed. This contribution has an opposite sign with respect to the Standard Model top-gluon loop contribution. The compensation between the two is therefore reduced and the relative NLO contribution is more important amounting to about 70\% in this scenario. 

In this illustrative scenario, for a fixed value of the reheating temperature, the corrections account for a shift of about 50 GeV in the gravitino mass, which corresponds to a shift of about 10\%. For a fixed gravitino mass of 450 about GeV, the reheating temperature needs to be multiplied by about a factor of two in order to still satisfy the Planck constraint.

Let us emphasize that in a situation where the stops and the gluino are closer in mass to the annihilating stau, the impact of the presented corrections is therefore expected to be reduced and similar to what has been observed in the analysis of our Scenario I.
\section{Conclusion}
\label{Sec:Conclusions}

We have discussed the impact of NLO SUSY-QCD corrections and the QED Sommerfeld enhancement on the cross section of stau-anti-stau annihilation into top quarks as well as their impact on the relic density in scenarios where this cross section is important. We have explored a scenario where the lightest neutralino is the lightest supersymmetric particle (LSP) and a dark matter candidate and the mass difference between the neutralino and the lighter stau is small which increases the importance of the stau annihilation. As the stau annihilations are also important in scenarios with gravitino dark matter, we have analyzed the impact of NLO corrections on the gravitino relic density in a typical scenario with gravitino dark matter.

We have analyzed different ways of defining the theoretical uncertainty. We have shown that the usual way of using the variation of the renormalization scale largely underestimates the theoretical uncertainty in this case. It is better estimated by looking at different renormalization schemes. We have shown that at leading order, the uncertainty of the relevant cross section is about 30\% which translates into an uncertainty of about 5\% on the relic density. This uncertainty largely reduces at next-to-leading order and we have demonstrated that at NLO the theoretical uncertainty is comparable with the experimental one.

We have demonstrated that in the studied cases the next-to-leading order corrections are important. They shift the region in the parameter space which corresponds to the experimentally determined relic density by more than the experimental uncertainty. Moreover, the theoretical uncertainty of the next-to-leading order prediction for the relic density remains below 1\% making the NLO prediction for the relic density very precise.

\acknowledgments
We would like to thank H.~Kim and D.~St\"ockinger for interesting discussions concerning neutralino-stau co-annihilation in constrained MSSM scenarios. The authors would like to thank A.~Pukhov for providing us with the possibility to interface our {\tt DM@NLO} code with {\tt micrOMEGAs}. J.\,H.\ is supported by the DFG Emmy Noether Grant No.\ HA 8555/1-1. The work of B.\,H.\ is supported by {\it Investissements d'avenir}, Labex ENIGMASS, contrat ANR-11-LABX-0012. The work of S.\,S.\ is supported by the DFG through RTG 2149 {\it Strong and Weak Interactions -- from Hadrons to Dark Matter}. The figures presented in this paper have been generated using {\tt Matplotlib} \cite{Matplotlib} as well as {\tt JaxoDraw} \cite{Jaxodraw2003, Jaxodraw2008}.

\bibliographystyle{apsrev}

\end{document}